\documentstyle[multicol,aps,epsfig]{revtex}
\protect{
\setlength{\columnwidth}{9.3cm}}
\def\3{\ss}
\def\La{L$_\alpha$}

\def\Pb{P$_{\beta '}$}
\def\H2{${\rm H_{I\! I}}$}
\def\C{$\;^\circ$C$\,$}

\begin{document}

\title{
Sterols sense swelling in lipid bilayers}
\author{
Frank Richter$^{1,2}$, Leonard Finegold$^3$ and Gert Rapp$^1$\cite{byline}}
\address{
$^1$ European Molecular Biology Laboratory,
Outstation Hamburg, DESY, Notkestra\3e 85, D-22603 Hamburg, Germany\\
$^2$ Department of Physics, E22 Biophysics, TU Munich, 
D-85748 Garching, Germany\\
$^3$ Department of Physics, Drexel University, Philadelphia, PA 19104, USA\\
}
\date{November 5, 1998}
\maketitle
\begin{abstract}
In the mimetic membrane system of phosphatidylcholine bilayers, 
thickening (``pre-critical behavior'', ``anomalous swelling'') of 
the bilayers is observed, in the 
vicinity of the main transition, which is non-linear with temperature. 
The sterols cholesterol and androsten are used as sensors 
in a time-resolved simultaneous small- and wide angle x-ray 
diffraction study to investigate the cause of the thickening. 
We observe precritical behavior in the pure lipid system, as well 
as with sterol concentrations less than 15$\%$. 
To describe the precritical behavior we introduce a theory of 
precritical phenomena.
The good temperature resolution of the data shows 
that a theory of the influence of fluctuations needs modification. 
The main cause of the critical behavior 
appears to be a changing hydration of the bilayer.
\end{abstract}
\begin{multicols}{2}
\tighten
\section{INTRODUCTION}

Our ultimate aim is a fuller description of biological cell membranes, 
via the physics of membrane models. Cell membranes are mainly in equilibrium, 
yet obviously must experience fluctuating non-equilibrium conditions 
during cellular events such as membrane fusion and transmembrane signaling. 
An excellent experimental model system for biological cell membranes is 
a dispersion of phospholipid in water, known as liposomes, which exhibit 
a variety of interesting phases as temperature is changed \cite{Sack95}. 
Major changes in physical properties are seen at the gel to fluid 
phase transition, the main transition at temperature $T_m$, which is a 
first-order transition \cite{Nag,Fine2}, 
occurring over a narrow temperature range. However, when approaching 
the main transition from above $T_m$, structure-sensitive methods reveal 
a nonlinear increase (with decreasing temperature) of the lattice 
parameter $d$ in phosphatidylcholine (PC)-membranes \cite{Zhang,Chen,Jesp},
which is attributed to critical behavior (illustrated in Fig. \ref{temps}). 
This increase is here termed ``anomalous'' behavior. This term has been
used also in other contexts \cite{Jesp}.
The corresponding 
critical temperature $T_c$ would lie at a temperature just below 
that of $T_m$, 
and so is ``hidden'' by the main transition. Critical behavior is also 
seen in liposomes by nuclear magnetic resonance (NMR) \cite{Hawton}
and ultrasonic absorption \cite{Mitaku}. This behavior in 
phosphatidylcholine membranes, a major component of cell membranes, 
is believed to be of physiological importance, because small 
temperature changes change the thickness of the membrane drastically 
and thus can easily affect protein functions \cite{Gruner,Bloom}. Additional 
support for biological relevance comes from the fact that the nonlinear 
behavior occurs in the neighborhood of a phase transition, a 
non-equilibrium state of matter.

The bilayer thickness $d$, measured by x-ray diffraction, is composed 
of three regions which are the headgroup thickness $d_H$, which is part 
of the bilayer thickness $d_B$, and the inter-bilayer water layer $d_W$, 
where $d_B + d_W = d$. The aim is to disentangle the contribution of 
each region to the observed critical thickening in d, by examining four 
physically different models, each of which affect mainly one region; 
the models do not necessarily agree \cite{Zhang,Chen,Jesp,Jesp2}. 
In brief, Model I (following the description of the models by 
Nagle \cite{Nag}) attributes the thickening to critical fluctuations, 
where a softening of the bilayer leads to a reduced bending modulus 
(via enhanced Helfrich undulations) \cite{Helf}, and affects 
mainly $d_W$ \cite{Lip,LipLei}. This model has been tested and supported
by neutron scattering analysis \cite{Jesp,Jesp2}. 
An analysis of x-ray data \cite{Zhang} did not confirm 
the reduced bending modulus. Model II attributes the thickening to
an increasing conformational order in the hydrocarbon 
chains upon approaching $T_c$, and hence to an increase in the lipid 
bilayer thickness $d_B$ \cite{Zhang}. Both analyses use mean field-type 
liquid crystal theories whose application to critical phenomena and 
phase transitions -- according to the Landau theory the 
\La-\Pb-transition has no 
chance of being second order \cite{deG} -- is debatable. In Model III, 
the thickening is attributed to interbilayer hydration and van der 
Waals forces, which increase the water layer thickness $d_W$ \cite{Nag}. 
Reasons for a change in these forces when approaching the transition 
and their effect on the bilayer are not well understood, nor is the 
functional form modeling these forces tested in the 
neighborhood of the main transition. Model IV attributes the thickening 
to an increased headgroup layer thickness $d_H$ only, which is, however,
not supported by NMR work \cite{Gaw}.

Sterols are an important constituent of eukaryotic membranes. In mammals, 
cholesterol is essential for the well-being of
cells \cite{Fine2}; a concentration-dependent effect on softening 
and stiffening the membrane has been found for cholesterol \cite{Jesp2}. 
Sterols can also play a role in some protein functions coupling lipid 
dynamics to protein dynamics \cite{Fine2}, and in the action of an 
acetylcholine receptor \cite{Rankin}. 
Cholesterol organizes submicron domains in living cells \cite{Varma98}.
Here, in the absence of proteins, 
sterols serve us as a natural sensor of the lipid dynamics. In this 
context it is worthwhile to mention that a high content of 
cholesterol ($>20$ mol$\%$) changes the lipid phase diagram drastically, 
whereas small amounts ($<10$ mol$\%$) do not affect the lipid phase diagram 
significantly \cite{Fine2}. Note that a phase diagram contains 
information about the equilibrium phases, and that it does not give 
direct information about non-equilibrium effects which may naturally 
occur at the phase boundaries. It has been shown \cite{Jesp2}
that small amounts of 
cholesterol strongly affect the structure of the lipid membrane near 
the main transition.

Our first aim in studying the phenomenon of increased swelling is to 
find out whether it is analytical or indeed anomalous. If the latter 
is found, i.e. the d-spacing close to a transition can be fit to 
an expression of the form 
$\left[(T - T_c)/T_c\right]^{-\alpha}$, where $T_c$ is the 
critical temperature and $\alpha$ the critical exponent, then we intend to 
quantify the effect by determining the critical exponent, and test the 
theoretical predictions. Critically behaving bilayer systems are of 
interest because of their low dimensionality and the implicit 
possibility of long range ordering effects in fluid phases.

The main goal is to study various membrane interactions in order to 
identify the source of the anomalous swelling in PC-membranes. We probe 
the interactions with two different sterols, cholesterol and androsten. 
In terms of structure, cholesterol resembles androsten prolonged by an 
extra aliphatic tail. The relative sizes and position of the two 
sterols and the PC bilayer are such that the sterols are about 
half the length of a lipid molecule, and they lie towards the water 
interface \cite{Fine2}. 

\section{EXPERIMENTAL}
\subsection{Protocol}
Using time-resolved simultaneous small- and wide angle x-ray diffraction 
(SAX and WAX), we are able to retrieve structural information
in the region about the main transition 
as well as in the neighboring phases. The experiments were conducted at 
beam line X13 of EMBL at DESY, Hamburg. Beam line \cite{Rapp92} and
data acquisition system \cite{Rapp} have been
been described in detail. Here, we report measurements 
on dimyristoylphosphatidylcholine (DMPC), though we have performed 
analogous experiments on dilauroylphosphatidylcholine (DLPC) and 
dipalmitoylphosphatidylcholine (DPPC). The DMPC and cholesterol were from
Avanti (Alabaster, AL), and androsten (androsten-3$\beta$-ol) from
Steraloids (Wilton, NH).
Sample preparation is described 
elsewhere \cite{Fine2}.
A single experiment consisted of a first heating from 4\C to 29\C, 
i.e. to more than 5\C above the main transition 
temperature. The sample was then cooled to 10\C, followed by a second 
heating up to 29\C; i.e. each cycle consisted of three ramps. This 
procedure was chosen to exclude effects from improper mixing of lipids 
and sterols during sample preparation, to single out phase effects caused 
by a slow kinetics and/or metastability, and to look at possible 
hysteretic behavior. The scan rate was 0.5 \C/min within the 
phases, and 0.25 \C/min between 19\C and 26\C covering the main transition. 
This rate was determined in test experiments with the lipid only.
A second sample stayed during the 
experiment in the same sample holder but was not exposed to the beam. It was 
always measured as a reference, prior to and after 
the temperature cycle, to check for possible radiation damage. 

\subsection{Data}
From our x-ray data we directly determine the positions of the first- 
and second-order lamellar reflections, as well as the wide-angle 
reflections due to the ordering of the aliphatic chains.
Typical simultaneous small- and wide-angle x-ray diffraction patterns 
of vesicles of DMPC with 10 mol$\%$ androsten
are shown in Fig. \ref{spec}. 

\begin{center}
\begin{minipage}{8.9cm}
\begin{figure}[htb]
\setlength{\unitlength}{1cm}
\begin{picture}(8.5,6.5)
\epsfig{file=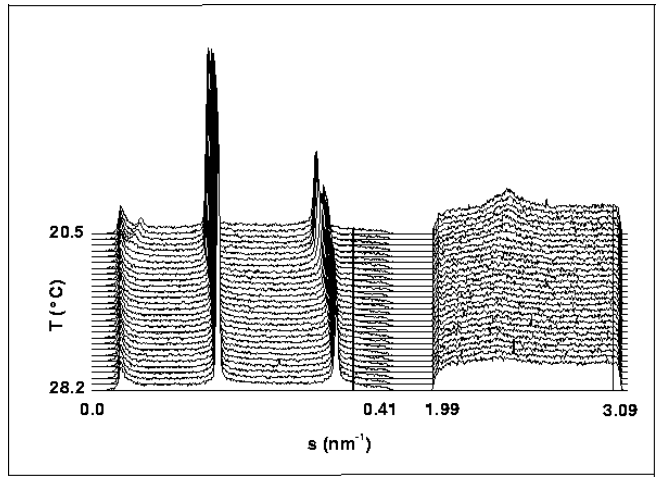,bbllx=550pt,bblly=130pt,bburx=1142pt,bbury=1794pt,width=8.5cm,angle=-90}
\end{picture}
\caption{
\label{spec}
Small and wide angle x-ray diffraction patterns of 
dimyristoylphosphatidylcholine
vesicles with 10 mol$\%$ androsten. Raw data of the first
heating through the main transition is shown. The scattering vector 
$s=1/d=(2\sin \vartheta )/\lambda )$, with $2\vartheta$ the Bragg angle.}
\end{figure}
\end{minipage}
\end{center}
\par
The swelling is seen by the movement of the
layer reflections in the SAX-region towards smaller $s$-values (i.e. smaller
angles) upon approaching the the main transition. The ripple phase is
identified by the (10)-reflection, that is seen near the beam stop edge,
as well as by the rising reflection from the ordering chains in the
WAX-region.

We recorded 
two orders of reflections in order to better determine the position 
of the reflections since, especially in the ripple phase and in the 
coexistence region, there is an overlap due to scattering from the 
in-plane ripple structure. 
Fig. \ref{dT} shows the d-spacing $d(T)$ for the variety of 
samples measured. 
\begin{minipage}{8.9cm}
\begin{figure}[htb]
\setlength{\unitlength}{1cm}
\begin{picture}(8.5,6.5)
\epsfig{file=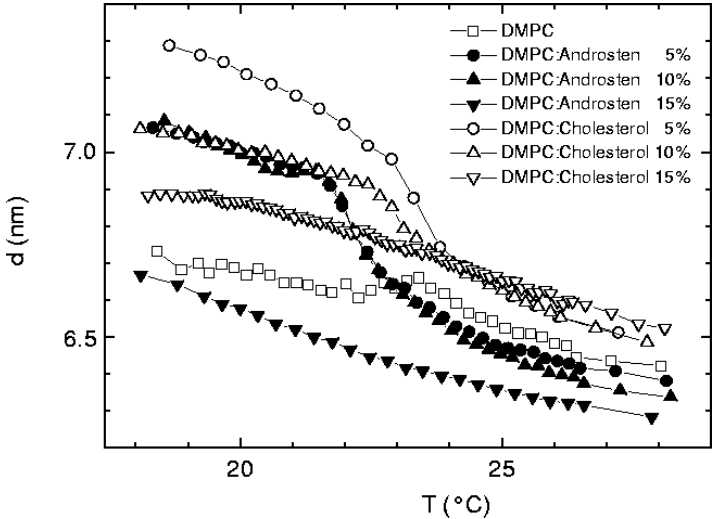,bbllx=550pt,bblly=130pt,bburx=1142pt,bbury=1794pt,width=8.5cm,angle=-90}
\end{picture}
\caption
{\label{dT}
The effects of cholesterol and of androsten
on the bilayer thickness $d$ of 
dimyristoylphosphatidylcholine vesicles:
results of 
the first heating (see text) are shown. The experimental points are connected
by lines to aid the eye.} 
\end{figure}
\end{minipage}
\section{RESULTS}
\subsection{Determination of $T_m$ from x-ray measurements}
If only SAX data are collected for measuring $T_m$, 
then for PC systems the transition 
temperature $T_m^{sax}$ might be determined as a major change in 
slope $d\; d(T)/d T$ at the low-temperature end of the swelling \cite{Nag}. 
In particular, when the d-spacing of the two phases participating in 
the transition is continuously changing,
$T_m$ can not be determined with precision. In combined time-resolved 
SAX and WAX studies we could define a transition 
temperature $T_m$ at which the intensity of the chain 
reflection recorded in the ripple phase is reduced to half of its 
maximum. Instead, taking into account the coexistence of the ripple 
phase and the \La-phase, we here define a transition zone $\Delta T_m^{wax}$
over which the intensity drops with increasing temperature.
Though the WAX measurements describe the chain melting process 
characterizing $T_m$ (and the degree of chain ordering could serve as order 
parameter), these measurements depend -- to a certain degree -- on long 
range periodic order among the chains. Moreover, it is by no means 
obvious that the interactions responsible for the chain ordering are 
directly related to the critical bilayer swelling $d$. One requires, in 
addition, a relation between the cooperativity of chain ordering and 
critical behavior.

The d-spacing as well as 
the intensity of the ripple reflection exhibits a very strong 
temperature dependence near the main transition, as already studied 
in detail \cite{Mat}, which is evidence of the non-equilibrium state 
of the ripple phase at least near the main transition. The formation of 
the ripple phase itself might thus be a critical phenomenon, too.
The existence of the ripple phase appears
to be linked to the processes on the higher temperature side of the main 
transition.
We also note that of all the samples being considered for anomalous 
behavior only those displaying a local maximum in $d(T)$ at the main 
transition show a clear (10)-ripple reflection that forms simultaneously 
during the main transition (Fig. \ref{cont}). 
\begin{minipage}{8.9cm}
\begin{figure}[htb]
\setlength{\unitlength}{1cm}
\begin{picture}(8.5,6.5)
\epsfig{file=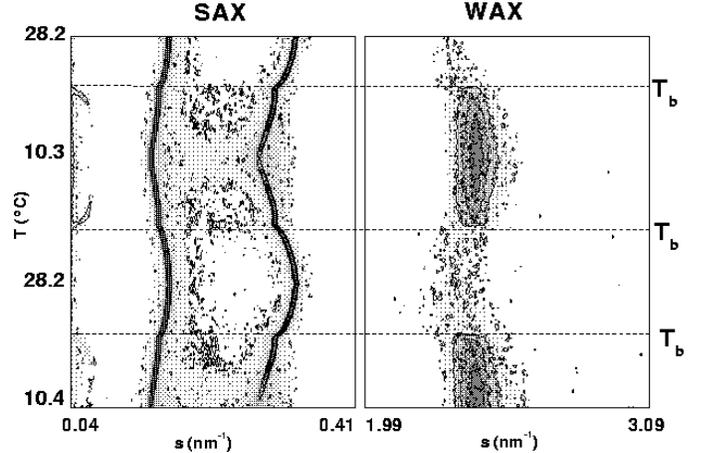,bbllx=550pt,bblly=130pt,bburx=1142pt,bbury=1794pt,width=8.5cm,angle=-90}
\end{picture}
\caption
{\label{cont}
Contour plot of the intensities recorded in the 10 mol$\%$ androsten 
(in dimyristoylphosphatidylcholine vesicles) experiment as function of 
the scattering vector $s$ and temperature $T$ beginning at 10.4\C.
Small- (first two orders seen) and wide-angle (chains) x-ray 
diffraction patterns are shown.
$T_b$ 
is defined in the text at Table \ref{data}.}
\end{figure}
\end{minipage}
\subsection{Is the swelling critical ?}
\subsubsection{Critical Behavior}
Any singular behavior (power-law- or logarithmic-divergent, as well 
as a finite peak), of a quantity $f(T)$ about a critical temperature $T_c$, 
can be embraced in one single functional formula \cite{Bin}:
\begin{eqnarray} 
f(T)&\sim&\frac{1}{\alpha}
\left(\left\vert\frac{(T-T_c)}{T_c}\right\vert^{-\alpha}-1\right)
\end{eqnarray}
where $\alpha$ is the critical exponent. We preserve the name $\alpha$ 
to point out the similarity of the lattice swelling to the 
increasing specific heat (with decreasing temperature) in the transition. 
In a log-log representation of $f(T)$, a fit then yields the 
critical parameters $\alpha$ and $T_c$.
If a nonlinear, yet well behaved, graph of the log-log 
data is obtained the function $f(T)$ can be described analytically. 
Therefore, the representation of formula (1) serves as a test to 
distinguish between analytical and critical behavior. A critical 
exponent of zero needs some further investigation, since it either 
indicates a logarithmic divergence, a cusp-like behavior, or a perfectly 
analytical function with no anomalous behavior worse than a jump 
discontinuity \cite{Stan}. The log-log-plots (Fig. \ref{loglog}) 
disclose a critical behavior of $d(T)$.

\begin{center}
\begin{minipage}{8.85cm}
\begin{figure}[htb]
\setlength{\unitlength}{1cm}
\begin{picture}(8.5,6.5)
\epsfig{file=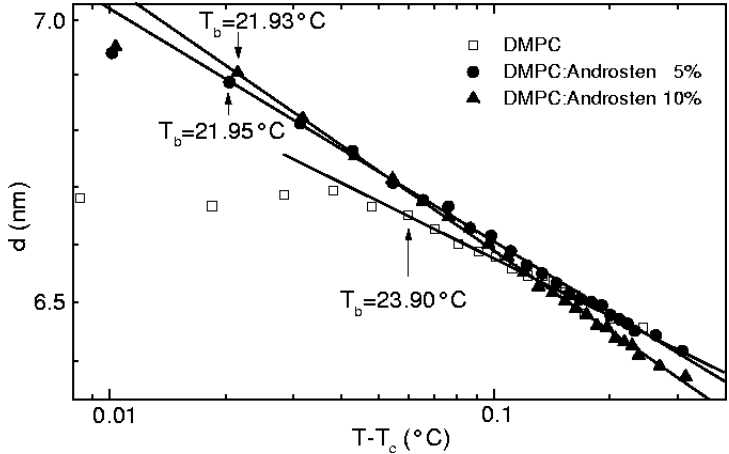,bbllx=550pt,bblly=120pt,bburx=1152pt,bbury=1794pt,width=8.5cm,angle=-90}
\end{picture}
\caption
{\label{loglog}
Anomalous swelling: log-log-plot of the bilayer
thickness $d$ (of vesicles of dimyristoylphosphatidylcholine with 0, 5 
or 10 mol$\%$ of androsten) is fitted to Expression (2).
Linear fits yield $\alpha =$
0.01175, 0.02715 and 0.03074 respectively
(first heating, see text and Table \ref{data}).
The last point on the line is labelled $T_b$.}
\end{figure}
\end{minipage}
\end{center}
\vspace{0.1cm}
\par
Here we address the problem of determining $T_c$ for a precritical 
system, i.e. where $T_c$ is hidden by the phase transition. 
Formally, this problem is 
implicit to the definition of a critical exponent (Formula 1) which implies 
the knowledge of $T_c$.
\subsubsection{Determination of T$_c$}
When a precritical system is in the swelling region where it displays 
critical behavior it does not ``know'' that it will, upon approaching $T_c$, 
undergo a first order transition. So we can treat the precritical system 
over the swelling region as a critical system \cite{Hawton,Mitaku}.
The fact that the system does not ``know'' {\it a priori} the order of the 
transition it will undergo when it swells critically is the base for 
the theory of critical unbinding of membranes \cite{LipLei}, which means the 
d-spacing can diverge. Therefore, we can reduce Formula (1) to 
the case $\alpha >0$ \cite{Bin}.
\begin{eqnarray}
\ln d(T)&\sim&\ln\left(\frac{(T-T_c)}{T_c}\right)\hspace{0.22cm}
{\rm or}\hspace{0.22cm}d(T)\sim\left(\frac{(T-T_c)}{T_c}\right)^{-\alpha} 
\hfill .
\end{eqnarray}
In the following, we use Fig. \ref{loglog} (DMPC, first heating) as an 
example.
To obtain $T_c$ we employ a least squares minimization technique while 
varying $T_c$. If the system behaves critically then all data points 
(starting from the highest temperature data 
point $T_i$) will follow a linear 
function in the log-log plot, upon appropriate choice of $T_c$. The 
advantage of this method is that, beginning at the highest 
temperatures $T_i$ (for this example $T_i=28.05$\C), it first finds the 
maximum number of data points $b$ that could follow a linear function, 
by comparing the measured d-spacings under the constraint 
$d(T_i)-d(T_{i+1}) < d(T_{i+1})-d(T_{i+2})$, $i=\{ 1,b\}$. 
The temperature recorded at the data point at the low-temperature 
end ($T_b=23.90$\C) of the thus-defined region of anomalous swelling 
gives the starting temperature of the $T_c$ variation. 
The system behaves critically if all the data points in the anomalous 
swelling region indeed make a linear function in the 
log-log-representation; if not, the swelling is not critical. 
In the next step, if the data follow a linear function, the slope of 
this linear function (i.e. the critical exponent) is found by 
minimizing the sum of squares of deviations from the line. 
All of the data points enter with the same weight. The first 
data point outside the purely critical regime (at 23.60\C) is easily 
distinguished, even by visual inspection, by its large deviation 
off the linear fit in the log-log representation (Fig. \ref{loglog}). 
The preceding non-deviating point, defining the ``breakdown temperature''
$T_b$, where the critical behavior breaks down,
is at 23.90\C in Fig. \ref{loglog}. The existence of such a data 
point follows from the observation of precritical phenomena. 
The procedure finds the global minimum because raising or lowering of 
$T_c$ yields a systematic concave or convex
distortion, respectively, of the string 
of data points from a straight line.

Table \ref{data} gives $T_c$, the temperature $T_b$ of 
the last non-deviating data point in the critical regime closest to $T_c$, 
and the critical exponent $\alpha$, along with the transition temperatures 
$T_m^{sax}$ obtained from the $d(T)$-plot,
and $\Delta T_m^{wax}$,
as well as the amount of anomalous swelling observed ($d(T_b)-d(28$\C)). 
These temperatures are illustrated in Fig. \ref{temps}.

\begin{center}
\begin{minipage}{8.85cm}
\begin{figure}[htb]
\setlength{\unitlength}{1cm}
\begin{picture}(8.5,6.5)
\epsfig{file=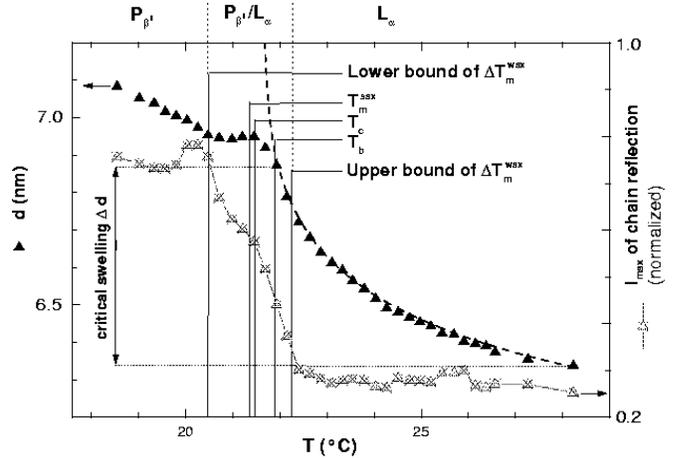,bbllx=550pt,bblly=115pt,bburx=1162pt,bbury=1794pt,width=8.5cm,angle=-90}
\end{picture}
\caption
{\label{temps}
Illustration of transition-related temperatures: lattice spacing $d$ versus
temperature (upper graph, closed symbols) 
measured for dimyristoylphosphatidylcholine
vesicles with 10 mol$\%$ androsten during the first heating. The dashed
line is a fit to Expression 2. The lower graph shows the
maximum intensity of the chain
reflection (normalized to the intensity measured at $T=10$\C). 
The drop in intensity indicates the melting of the chain
lattice during the main transition.}
\end{figure}
\end{minipage}
\end{center}
\vspace{0.1cm}
\par
The critical behavior is observed (Fig. \ref{loglog}) in the region 
closest to the 
critical point over at least a full decade in $(T - T_c)/T_c$, where data 
have been collected.
For all of the samples listed, the single value of $T_c$ itself is to be 
interpreted with caution. We remind ourselves that we do not directly 
look at the critical behavior about a second order transition point $T_c$. 
Instead, the first order transition intercedes, and thus eliminates 
from observation the region in the immediate vicinity of $T_c$. 
This region is the 
most important for determining the critical exponent for critical 
phenomena, because there the accompanying long-wavelength 
fluctuations govern the physical properties of a critical 
system and have to be taken into account, e.g. by means of 
RG-theory \cite{HH}.
\end{multicols}
\begin{table}[htb]
\caption{
Critical parameters, $T_c$ (critical temperature), 
$T_b$ (breakdown temperature) and $\alpha$ (exponent), for
dimyristoylphosphatidylcholine vesicles with two sterols at various
concentrations (conc.). Also given are the main transition temperatures
from small- and wide-angle x-ray diffraction
$T_m^{sax}$, $\Delta T_m^{wax}$, and the anomalous swelling $d(T_b)-
d(28$\C)(illustrated in Fig. \ref{temps}).}
\par\noindent
\begin{tabular}{lrccccccc}
$Sterol$&$Conc.$&$T_m^{sax}$&$\Delta T_m^{wax}$&$T_c$&$T_b$&$T_b-T_c$&$\alpha$&
$d(T_b)-d(28$\C)\\
&mol$\%$&\C&\C&\C&\C&\C&&nm\\ 
\tableline
$1^{st}\, Heating:$&&&&&&&& \\
\\
None&0&23.3&23.3-23.9&22.55&23.9&1.4&0.02175&0.24\\ 
Androsten&5&21.2&21.0-22.3&21.51&21.9&0.4&0.02715&0.47\\
Androsten&10&21.3&20.5-22.2&21.47&21.9&0.4&0.03074&0.54\\
Cholesterol&5&22.6&&23.03&23.3&0.3&0.01771&0.39\\
Cholesterol&10&22.4&21.9-23.0&21.24&22.4&1.2&0.03827&0.42\\
\\ 
$1^{st}\, Cooling:$&&&&&&&& \\
\\
None&0&23.0&23.6-22.8&22.45&23.7&1.3&0.02512&0.27\\
Androsten&5&21.0&22.0-19.9&20.90&22.1&1.2&0.03796&0.42\\
Androsten&10&21.0&21.8-21.0&21.06&22.1&1.0&0.03409&0.39\\
Cholesterol&5&22.2&&21.86&22.5&0.6&0.02714&0.44\\
Cholesterol&10&22.1&22.4-22.0&19.72&22.2&2.5&0.05164&0.40\\
\\
$2^{nd}\, Heating:$&&&&&&&& \\                          
\\
None&0&23.2&23.2-23.7&22.89&23.9&1.0&0.02264&0.30\\
Androsten&5&&20.0-22.0&21.28&22.1&0.8&0.03445&0.46\\
Androsten&10&21.3&21.2-22.3&21.40&21.9&0.5&0.03329&0.55\\
Cholesterol&5&23.0&22.8-23.4&22.72&23.0&0.3&0.02453&0.40\\
Cholesterol&10&22.3&21.8-22.9&20.90&22.7&1.8&0.04173&0.37\\
\end{tabular}
\label{data}
\end{table}
\par
\begin{multicols}{2}
The advantages of our approach to investigate the criticality of a system
are manifold:
no {\it ab initio} model needs to be assumed, which 
is a valuable asset especially for critical 
phenomena and phase transitions; no fitting of experimental data other 
than the linear fit (to the log-log plots) to determine the 
exponent is needed; and no adjustable parameters are used. The 
breakdown temperature $T_b$, replacing the non-observable critical 
temperature $T_c$ when one applies the theory of critical phenomena 
is not very sensitive to the fit. On the other hand, $T_c$ and $\alpha$ 
are more sensitive to the fit, also due to the relatively small
temperature range where data have been collected. However, the
exact value of the critical temperature $T_c$, and hence the
obtained value of the exponent, is not crucial for the interpretation 
of precritical phenomena. We stress that it is the breakdown 
temperature $T_b$, and not the critical temperature $T_c$, that is 
characteristic for precritical phenomena. For critical phenomena $T_b=T_c$.
In our experiments we find in all cases 
$T_b>T_c$.
\subsubsection{The precritical parameters}
The exponents derived are all close to zero, and in the light of the 
theory of critical phenomena, a small magnitude of an exponent implies 
a sharp divergence \cite{Stan}. The exponent $\alpha$ is the quantity 
that allows us to rank precritical phenomena.
With one exception $\alpha$ is always smaller for the pure lipid
than for the sterol samples (Table \ref{data}). 
Following the theory for critical phenomena, this implies that the 
$d(T)$ dependence for the pure lipid is the closest to criticality, 
i.e. the pure lipid would display a singularity of a relatively 
higher order at $T_c$ than would the sterol samples. Indeed, looking 
at the $d(T)$-plots obtained from the layer reflections (Fig. \ref{dT}), the 
pure lipid shows the flattest region between 28\C and $T_m^{sax}$ 
(at {\it ca.} 23.5\C). We conclude that the higher the concentration of 
sterol (for those samples that display critical behavior) the higher 
is $\alpha$, i.e. the less critical is the behavior. This is related 
to the breakdown of criticality further away from $T_c$.

It is further noticed, for samples showing the strongest swelling 
effect in Fig. \ref{dT} (i.e. $d(T_b) - d(28$\C) for 5 mol$\%$ 
cholesterol and 
10 mol$\%$ androsten), that the two temperatures obtained in our analysis 
of the precritical behavior ($T_c$ and $T_b$) are relatively close 
(Table \ref{data}). 
Then, both temperatures have a qualitatively similar dependence on the 
sterol concentration pointing towards a relationship among each 
other.
\subsection{Transition-related temperature correlations}
Table \ref{data} shows that the lowering of $T_m^{sax}$ depends on the sterol,
since androsten lowers $T_m^{sax}$ more than cholesterol. 
For both sterols 
the two concentrations tested, $T_m^{sax}$ and 
$\Delta T_m^{wax}$ do not 
differ significantly. However, the number of concentrations measured is
not enough to conclude that these temperatures are only 
weakly dependent on the sterol concentration, as long as the system 
shows the anomalous swelling \cite{Jesp2}. There is only a small 
temperature hysteresis. For $T_c$ we see a similar behavior with the 
exception that there is an obvious dependence on the cholesterol 
concentration. The dependence of $T_b$ on the sterol concentration is 
analogous to that of $T_m^{sax}$, but the lowering is not as pronounced. 
$\Delta T_m^{wax}$ features the same tendency as $T_b$.

The samples displaying the largest swelling effect (i.e. 
$d(T_b) - d(28$\C)), 
which are the 5 mol$\%$ cholesterol and
the 5 and 10 mol$\%$ androsten samples 
(see Table \ref{data}), 
yield a lower $T_m^{sax}$ than $T_c$. Of course, this does not 
mean that these samples show a real continuous transition with an 
observable critical point, since we always find $T_m^{sax}<T_b$, i.e. on 
cooling, the swelling remains precritical and the system never 
reaches the maximum swelling before it leaves the critical regime. 
On the other hand the breakdown of critical behavior
at sub-maximal swelling (e.g. for 5 mol$\%$ androsten 
$d(T_b)=6.86$ nm$ <d(T_m^{sax}=6.95$ nm)
(cf. Fig. \ref{temps}) implies that 
the systems continue to swell below $T_b$ but not in a critical manner. 
This non-critical swelling, accompanied by the continued chain ordering 
($T_b$ is always larger than both $T_m^{sax}$ and the lower end 
of the transition zone $\Delta T_m^{wax}$ (Table \ref{data})), is evidence 
that the chain ordering does not obey critical dynamics in the immediate 
neighborhood of the maximum swelling. On heating a 
decreasing chain ordering, combined with a noncritical reduction of 
the lattice parameter in the vicinity of its maximum, is seen. $T_b$ is 
again always larger than $T_m^{sax}$ and the lower 
end of the transition zone $\Delta T_m^{wax}$ (Table \ref{data}). 
Therefore, only at 
temperatures $T_b$ higher than $T_m^{sax}$ does the system move 
into the critical 
regime (Table \ref{data}). 
For all samples, we generally find $T_m^{sax}$ on the 
lower temperature side of the transition zone $\Delta T_m^{wax}$ 
(Table \ref{data}). 
Thus, on heating the d-spacing does not start to decrease before there 
is a decrease in the chain ordering. On cooling, the swelling does not stop
before there is already periodic chain ordering established. 
This relationship between the chain ordering and the entire swelling --
critical and non-critical -- seems to be independent of the 
sterol used and also of its concentration.
\subsection{Effect of sterols}
\subsubsection{Membrane swelling}
We see (Table \ref{data}) 
that the more pronounced a sample shows the swelling 
effect (i.e. $d(T_b) - d(28$\C)) the more $T_c$ moves towards 
the high temperature end of the transition region $\Delta T_m^{wax}$. 
The 10 mol$\%$ 
cholesterol appears to not match this rule because
the sterol concentration is obviously high enough to diminish
the critical swelling \cite{Jesp2}.
Only the androsten samples actually have a $T_c$ inside the 
transition region $\Delta T_m^{wax}$. Since the sterols, as basically 
hydrophobic entities incorporated into the lipid bilayer, interact 
in the first place with the aliphatic chains, they decrease 
the tilt angle of the ripple phase \cite{Need} upon incorporation. 
As a consequence they also increase the headgroup hydration 
by increasing the headgroup-headgroup spacing \cite{Jesp2}, accompanied by 
(critical) fluctuations resulting a possible softening of the 
bilayer which in turn could cause an increased water layer 
(cf. Model I). One might argue that the sterols overpower 
the origin of the anomalous swelling, a PC property, 
by decreasing the chain and/or headgroup tilt. This is not 
the case since the swelling remains critical for smaller amounts of 
sterols as shown above. A measure of the impaired periodic chain ordering 
(due to the presence of the 
sterols)
is the width of the chain reflection dependent on the 
sterol concentration during the main transition. The 
chain reflection widths (measured in detector channels) are: 
49 (no sterol), 52 (5 mol$\%$ androsten), 58 (10 mol$\%$ androsten), 
56 (5 mol$\%$ cholesterol) and 63 (10 mol$\%$ cholesterol). 
Because the chains 
are bound to the headgroups, the disorder induced in the chains 
should be transferred to the headgroups as well. Apparently, 
the impaired chain ordering, due to the sterols, yields a larger 
exponent $\alpha$ indicating a weakened criticality 
(Table \ref{data}) but is not a 
monotonous function of the sterol concentration. In other words,
the impaired
periodic chain ordering (width) mirrors the degree of criticality 
$\alpha$, but is 
not the origin of the anomalous swelling. 
\subsubsection{Cholesterol versus androsten}
For the cholesterol samples our data do not confirm the 
mutual exclusion of critical behavior and chain ordering because 
$T_b$ falls into the transition range $\Delta T_m^{wax}$. 
However, the tentative 
statement made above - that the chain ordering is not the origin of 
the critical behavior - remains valid, last but not least because 
increased chain ordering does break the criticality in the cholesterol 
samples, too, as evidenced by a $T_b$ within the transition range 
$\Delta T_m^{wax}$ (cf. Table \ref{data}). 
Thus, swelling in the presence of cholesterol 
can exist with limited periodicity in chain ordering. 
Possible explanations for the differing amount of chain ordering, 
required to break the critical behavior observed between the two 
sterols are (a) a sterol-specific interaction with the lipid 
molecules, and (b) an active domain-boundary seeking of sterols 
in the coexistence region.
Sterols, as sensors, have revealed an imperfect coupling of the chain 
ordering processes with the layer dynamics, i.e. the swelling, pointing 
towards a phenomenon affecting the hydrophilic-hydrophobic interface. 
Since none of the four models excludes changes in the headgroup-headgroup 
interaction, we now discuss the effects of 
critical fluctuations and hydration on these interactions.
\section{DISCUSSION}
\subsection{Testing the four Models}
In the remainder of this paper, we associate the experiments with the four 
models discussed earlier \cite{Nag}. Because undulations and fluctuations
are basic to Model I, we first discuss critical fluctuations and the
correlation of fluctuations with swelling. The effect of chain ordering on
the breakdown of criticality is connected with Model II, and hydration is 
connected with Models III and IV. The advent of a ripple phase appears to
be connected with the main transition and pre-criticality. Finally, 
thermodynamic and nonequilibrium effects are discussed, and the importance
of low scan rates is stressed.
\subsection{Critical fluctuations}
\subsubsection{Fluctuations in membranes}
Lipowsky and Leibler \cite{LipLei} predict, on the basis of Helfrich 
undulations, that the bilayer thickness d should be linear in $1/(T-T_c)$ 
(Model I), i.e. $\alpha = 1$. 
Our results (Table \ref{data}) give exponents $\alpha$ that differ 
significantly from 1, whereas earlier experiments on lipid 
membranes \cite{Jesp} and other soft condensed matter systems 
\cite{Safinya} supported the predictions \cite{LipLei}.
The magnitude of our precritical exponent $\alpha$ is reasonable, since 
an exponent of the order of 1 would yield a much larger swelling effect 
over the entire experimental critical regime of 28\C down to $T_b$.
Experimentally, even the largest amounts of critical swelling 
observed (Table \ref{data}, 
$d(T_b) - d(28$\C)) do not reach a tenth of the unit
cell size $d$ (of around 7 nm). When plotting our data (Fig. \ref{llip}) 
against $1/(T-T_c)$, we observe clear deviations from the predicted 
linear behavior \cite{LipLei}; a single linear fit is not reasonable.
\begin{minipage}{8.7cm}
\begin{center}
\begin{figure}[htb]
\setlength{\unitlength}{1cm}
\begin{picture}(8.5,6.5)
\epsfig{file=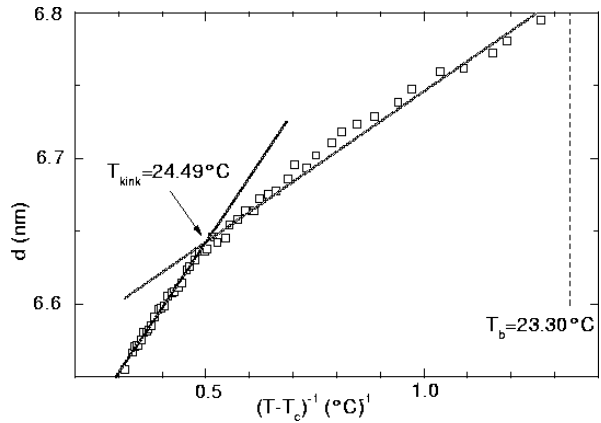,bbllx=535pt,bblly=100pt,bburx=1122pt,bbury=2154pt,width=8.5cm,angle=-90}
\end{picture}
\caption
{\label{llip}
Critical bilayer thickness (of vesicles of 
di\-myristoyl\-phos\-phati\-dyl\-choline,
first cooling, scan rate 0.1\C/min, $T_c=22.55\pm 0.05$\C);
a prediction \protect\cite{Lip} gives a straight line.}
\end{figure}
\end{center}
\end{minipage}
\par
Thermal fluctuations trigger the main 
transition. To investigate 
the role of enhanced fluctuations (all Models) on the swelling we 
consider the transition-related temperature derived from the layer 
reflections $T_m^{sax}$, since it marks the temperature at the end of 
the swelling, critical plus non-critical. Our data (Table \ref{data}) show 
that the addition of sterol shifts this transition temperature $T_m^{sax}$. 
This means, only a partial periodic chain ordering is required to 
obtain the maximum d-spacing in the transition region. Even more, 
for the samples showing the strongest swelling effect yielding 
a local maximum in $d(T)$ (Fig. \ref{dT}, 5 and 10 mol$\%$ androsten), 
the continued 
ordering upon lowering temperature reduces $d(T)$ slightly. 
A more pronounced local maximum in the transition range 
is seen in the pure DMPC 
as well, but here it covers a broader temperature range compared 
to the two sterol samples. This peculiar relationship between 
chain ordering and d-spacing leading to the local maximum might be 
explained by the changing contributions of in-plane and out-of-plane 
fluctuations. Whereas the in-plane fluctuations affect the periodic 
ordering of the chains, the out-of-plane-fluctuations in first 
place deteriorate the periodicity of the layer order. The lower 
energetic in-plane fluctuations are expected to contribute more at 
the lower temperature end of the swelling when the interlayer 
spacing becomes larger and larger and neighboring membranes are 
basically decoupled. However, when the chains start to order due 
to increasing chain-chain-coupling these fluctuations will then
be suppressed. In the theory of liquid crystals -lipid model 
membranes can be considered as such, in-plane fluctuations affect 
the bending modulus $K_c$ and the out-of-plane fluctuations affect the 
compression modulus $B$ \cite{deG}. Here, fluctuations broaden the 
width of the x-ray reflections as a function of the order caused 
by the decreasing correlation length of the scattering units. 
Fluctuations, or defects of the first kind, also reduce the maximum 
intensity with increasing diffraction order \cite{Gui}.
\subsubsection{Fluctuations: Correlation with swelling}
\vspace{-0.2cm}
The intensities of the x-ray reflections give information 
about membrane fluctuations. 
Our data show (Fig. \ref{ratio}(a)) a decrease in
the ratio of the maximum intensity of the first order to the second 
order reflections, with increasing sterol concentration and 
with decreasing temperature, in the swelling region (of 28\C down to $T_b$). 

Similarly, Fig. \ref{ratio}(b) shows that the ratios of the respective
integral widths 
(integrated inten\-sity/maxi\-mum intensity) has approximately
the same variation in $T$ for all sterol concentrations for $T>T_b$.

\begin{center}
\begin{minipage}{8.9cm}
\begin{figure}[htb]
\setlength{\unitlength}{1cm}
\begin{picture}(8.5,6.5)
\epsfig{file=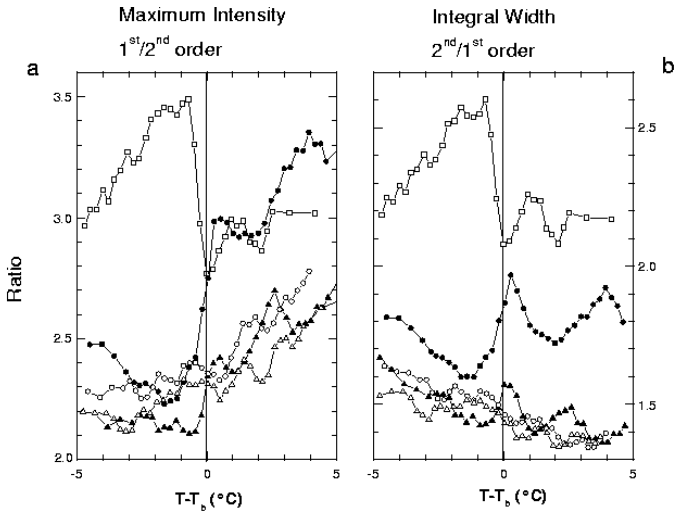,bbllx=550pt,bblly=110pt,bburx=1142pt,bbury=1794pt,width=8.5cm,angle=-90}
\end{picture}
\vspace{0.2cm}
\caption
{\label{ratio}Ratio of maximum intensities (a) and integral widths (b) of
first and second order reflections as function of temperature $T$ ($T_b$:
breakdown temperature) for vesicles containing up to 10 mol$\%$ sterol; 
symbols as in Fig. \ref{dT}. }
\end{figure}
\end{minipage}
\end{center}
\vspace{0.1cm}
\par
In light 
of the theory of x-ray diffraction, the similarity of the dependence of
both integral width
and maximum intensity ratios on
sterol concentration is to be interpreted as an increase of 
static defects \cite{Gui}, whereas the decreasing ratio of the 
maximum intensities with decreasing temperature
when approaching $T_b$ (i.e. with increasing swelling)
indicates a reduction of fluctuations (akin to a reduction in the 
Debye-Waller factor).
Since the reduction is 
relatively stronger in the sterol samples than in the pure DMPC, it 
follows that sterols reduce fluctuations.
In consequence we have to exclude a significant 
fluctuation-driven contribution to the critical swelling, i.e. Model I is
not supported. (However, 
they increase the criticality of the system as expressed by $\alpha$ 
(Table \ref{data})).
The observed
fluctuations are therefore not critical fluctuations and can be 
described in terms of the classical Ornstein-Zernicke theory 
\cite{LL}; in other words, the system is sufficiently far away from a 
critical point that a Landau-type theory is appropriate for the
description of the fluctuations. 

\subsection{Effect of chain ordering on the breakdown}
When comparing the breakdown temperature
$T_b$ to the main transition zone $\Delta T_m^{wax}$ 
(Table \ref{data}) we observe that 
$T_b$ virtually coincides with the upper bound of $\Delta T_m^{wax}$
for the pure lipid and 
the androsten samples on heating and cooling. For the cholesterol 
samples this is clearly not so (Table \ref{data}):
A certain degree of chain ordering appears to be permitted 
as $T_b$ moves into the transition zone $\Delta T_m^{wax}$. This 
ordering effect must be
related to the difference between the two sterols and their interaction with
DMPC. The fact that for the pure lipid $T_b$ coincides with 
the upper bond of periodic chain ordering (given by $\Delta T_m^{wax}$) is
persuasive evidence for the following scenario. On the one hand, $T_b$
marks (by definition) the observed lower temperature limit of the critical
behavior (as defined by Formula 2) in the \La-phase. On the other hand, the
upper bound of $\Delta T_m^{wax}$ denotes the higher temperature limit of the
periodic chain ordering of the ripple-\Pb-phase. Hence, on cooling, the onset
of chain ordering breaks criticality (as is illustrated in Fig. \ref{cont}
where the $T_b$ line intersects with the edge of a detectable WAX reflection.)
Therefore, changes in interactions within 
the hydrophobic membrane core, i.e. the steric repulsion and the 
van der Waals interaction between them, coupled to the chain ordering 
can not directly control the critical swelling, so Model II is not supported. 
(Models I, III and IV are indifferent towards chain ordering effects.)
Neither the steric repulsion 
nor the van der Waals interaction among the chains are expected 
to change drastically when entering the coexistence region, 
because both do not depend on the degree of periodic order. 
This is supported by the fact that one only observes critical 
swelling in PC but not e.g. in phosphatidylethanolamine (PE) membranes. 
(Additional data taken on mono- and dimethylated PE do show the swelling.) 

\subsection{Hydration}
The magnitude of the critical swelling $d(T_b) - d(28$\C)
is increased in the samples
containing a small amount of sterol ($\le 10$ mol$\%$) 
(Fig. \ref{dT} and Table \ref{data}), compared to that of the pure lipid. The 
adding of small amounts of sterol does not significantly decrease
the area per headgroup \cite{Zucker}, so that the difference in area
per molecule in the neighboring phases remains large 
(about $15\%-20\%$ \cite{Sack95}).
In turn
we expect an increased amount of water in the headgroup region 
possibly leading to a change in headgroup conformation, as is 
proposed by Model IV. NMR measurements indicate that water molecules 
might penetrate past the hydrophilic headgroup towards the hydrophobic 
membrane core \cite{Gaw}. The changing hydration affects the 
hydrophilic/hydrophobic interface. The dynamics of an interface is 
known to depend on the local curvature of the interface \cite{Tan} and 
can behave critically under certain conditions \cite{Meakin}. 
The addition of 15$\%$ sterol does not provide appropriate conditions to
observe critical swelling: Due to phase separation effects expected
(a miscibility gap \cite{Sack95} and the relatively low temperature in
these experiments, no critical behavior can develop (cf. Fig. \ref{dT}).
Also, the diffraction technique is insensitive to short-range order. We
have studies in progress on the miscibility gap \cite{Rapp99}. 
A change of the hydration shell about the headgroup is thus a 
candidate for the critical swelling effect. We further remark that a 
decreased headgroup area upon adding sterol does increase the amount 
of swelling but reduces the criticality of the swelling (Table \ref{data}). 
One is tempted 
to identify the kink in our results (Fig. \ref{llip}), at around 24.55\C, 
as the boundary of a critical hydration dominance. (The changes in slope in
Fig. \ref{llip} may be interpreted as a critical slowing down \cite{slow}.)
Lipowsky and Leibler 
imposed a hard confinement scenario in their derivation \cite{LipLei}, 
which is
perhaps not valid here \cite{Pet}, where a change in environment 
can be attributed to a change in hydration. Also, NMR data
\cite{Gaw} indicate a change in hydration at about this
temperature. Moreover, dynamical scaling can break down despite the 
persistence of interfacial undulations \cite{Sch}. To clarify the 
difference between the critical behavior of a changing headgroup 
hydration (Model IV) and bilayer thickening (Model II), we point out 
that the hydration depends on the headgroup environment
whereas the bilayer thickening 
in Model II is only
a property of the chain conformation. The headgroup environment and the 
hydrophobic chains are coupled, and thus headgroup and hydrocarbon 
chains could be considered as an integral entity (Model II). The 
differences between the two models
are (a) the coupling is not stiff, so the behavior of 
chains and heads can be different and might not be mapped directly 
onto each other, and (b) chain-chain interactions 
involve lipids only, whereas hydration dynamics requires solvent 
(here water) and lipid molecules. These differences could give a clue why a 
lipid system can possess only precritical behavior, but not 
critical behavior. 

When entering the coexistence region $\Delta T_m^{wax}$ the headgroups feel 
the ordering of the coupled chains. As shown above the chain ordering 
correlates with the breakdown of the critical swelling and hence breaks 
a critical interface dynamics. In response the system finds a new 
(metastable) equilibrium position forming the ripple structure. 
It has been suggested that
the swelling of the hydration shell at a hydrophobic/hydrophilic 
interface is a quite general phenomenon \cite{Svergun}. 
Also, osmotic pressure experiments on various PCs seem to show a 
critical dependence of the d-spacing on the osmotic pressure \cite{Pet}.

\section{SUMMARY AND CONCLUSION}

We demonstrated that the anomalous swelling in 
phosphatidylcholine membranes, of 14 carbons chain length, 
is precritical. By analyzing the nonlinear $d(T)$ behavior -- in terms 
of the theory of critical phenomena -- observed for the pure lipid 
and various sterol concentrations in the neighborhood of $T_c$, we were 
able to link the anomalous swelling to a changing headgroup hydration.

We propose that the formation of the ripple phase is the counterpart 
on the low-temperature side of the transition to the anomalous 
swelling on the high-temperature side. It is suggested
that the ripple phase and the
anomalous swelling are caused by the
hydration dynamics of the headgroups.

We have found no evidence of critical entropic bilayer undulations 
quoted in Model I. Model II suggests a critical thickening of the bilayer. 
The increase in the conformational ordering of the chains in the fluid 
phase, as they approach the coexistence region, is not known to obey 
critical dynamics.

Our sterol experiments show that adding small amounts 
of sterol does not break -- but reduces -- the criticality of the system, 
which seems to be linked to the larger headgroup area at the interface. 
This seems to be in agreement to Model III, suggesting that a 
decreased interfacial surface increases the hydration force \cite{McIntosh}, 
however a decrease in interfacial area over the region of anomalous 
swelling can not be supported by our data. This suggests that the 
hydration force does not change in a critical manner. The mechanism 
proposed by Model IV, a conformational change of the headgroups due 
to changing hydration, is supported by our experiments and can 
explain the critical behavior. However, the model does not explain 
the amount of swelling \cite{Nag}.

From our data the following picture emerges: the changing hydration 
of the headgroups is responsible for the critical behavior and 
indirectly coupled chain ordering dynamics provides the full magnitude 
of the swelling. Further, the similarity of the transition-related 
temperatures determined from the SAX- and WAX-reflections (Table \ref{data}) 
are evidence for a relationship of the conformational chain dynamics 
to the systems' transition from the fluid to the ripple phase, but 
not for the transition from the critical to the coexistence region, 
which is found to be related to the long range periodic chain 
ordering dynamics.

Whereas DMPCholine 
exhibits critical swelling, DMPEthanolamine does not \cite{Rappolt}. 
The much larger polar choline headgroup, the only structural difference
between both lipids, is easy to hydrate.
The existence of 
lattice critical swelling may well
be connected with this difference in changing the hydration, and
with the differences
observed, between DMPC and DMPE, in differential scanning calorimetry
\cite{Blume} of the main transition.

We emphasize that our 
precritical exponents describe the swelling behavior 
over the entire temperature range (where swelling is observed, in all
of our samples) supporting the subtle interface dynamics picture. 
However, a necessary requirement for an {\it ab initio} theoretical 
description is the inclusion of the formation of the ripple phase.
This work suggests that the critical behavior is caused by the hydration
dynamics.

\acknowledgements

L.F. and G.R. thank NATO for grant CRG 970225. L.F. is indebted to Michael
A. Singer, of Queen's University, for discussions.

\end{multicols}

\begin{references}
\bibitem[*]{byline}E-mail: frichter@embl-hamburg.de\newline
\hspace{1.5cm}L@drexel.edu\newline
\hspace{1.5cm}rapp@embl-hamburg.de
\bibitem{Sack95}E. Sackmann, in {\it Structure and Dynamics of Membranes},
edited by R. Lipowsky and E. Sackmann (North Holland, Amsterdam, 1995), ch.\ 5.
\bibitem{Nag}J.F. Nagle, in {\it Phase Transitions in Complex Fluids}, 
edited by P. Toledano and A.M.F. Neto (World Scientific, Singapore, 1998).
\bibitem{Fine2}L. Finegold and M.A. Singer, in {\it Cholesterol in 
Membrane Models}, edited by L. Finegold (CRC Press, Boca Raton, 1993).
\bibitem{Zhang}R. Zhang, W. Sun, S. Tristram-Nagle, R.L. Headrick, 
R.M. Suter, and J.F. Nagle, Phys.\ Rev.\ Lett. {\bf 74}, 2832 (1995).
\bibitem{Chen}F.Y. Chen, W.C. Hung, and H.W. Huang, Phys.\ Rev.\ Lett. 
{\bf 79}, 4026 (1997). 
\bibitem{Jesp}J. Lemmich, K. Mortensen, J.H. Ipsen, T. H{\o}nger, 
R. Bauer, and O.G. Mouritsen, Phys.\ Rev.\ Lett. {\bf 75}, 3958 (1995).
\bibitem{Hawton}M.H. Hawton and J.W. Doane, Biophys. J. 
{\bf 52}, 401 (1987).
\bibitem{Mitaku}S. Mitaku, T. Jippo, and R. Kataoka, 
Biophys. J. {\bf 42}, 137 (1983).
\bibitem{Gruner}S.M. Gruner and E. Shyamsunder, Ann. N.Y.\ Acad.\ Sci.\ 
U.\ S.\ A. 
{\bf 625}, 685 (1991).
\bibitem{Bloom}M. Bloom, E. Evans, and O.G. Mouritsen, 
Q. Rev.\ Biophys. {\bf 24}, 293 (1991).
\bibitem{Jesp2}J. Lemmich, K. Mortensen, J.H. Ipsen, T. H{\o}nger, 
R. Bauer, and O.G. Mouritsen, Eur.\ Biophys. J. {\bf 25}, 293 (1997).
\bibitem{Helf}W. Helfrich, J. Phys.\ (Paris) {\bf 47}, 321 (1986);
Z. Naturforschung {\bf 33A}, 305 (1978).
\bibitem{Lip}R. Lipowsky, Europhys. Lett. {\bf 7}, 255 (1988).
\bibitem{LipLei}R. Lipowsky and S. Leibler, Phys.\ Rev.\ Lett. 
{\bf 56}, 2541 (1986).
\bibitem{deG}P.G. de Gennes, {\it The Physics of Liquid Crystals} 
(Clarendon Press, Oxford, 1974).
\bibitem{Gaw}K. Gawrisch, cited in \cite{Nag}.
\bibitem{Rankin}S.E. Rankin, G.H. Addona, M.A. Kloczewiak, B. Bugge 
and K.W. Miller, Biophys. J. {\bf 73}, 2446 (1997).
\bibitem{Varma98}R. Varma and S. Mayor, Nature {\bf 394}, 798 (1998).
\bibitem{Rapp92}G. Rapp, Acta Phys. Pol. A {\ 82}, 103 (1992).
\bibitem{Rapp}G. Rapp, A. Gabriel, M. Dosiere, and M.H.J. Koch, 1995,
Nucl.\ Instr.\ Meth. A {\bf 357}, 178 (1995).
\bibitem{Mat}S. Matuoka, S. Kato, and I. Hatta, 
Biophys. J. {\bf 67}, 728 (1994).
\bibitem{Bin}J.J. Binney, N.J. Dorwick, A.J. Fisher, and M.E.J. Newman, 
{\it Theory of Critical Phenomena}
(Clarendon Press, Oxford, 1992), p.\ 5.
\bibitem{Stan}H.E. Stanley, {\it Introduction to Phase Transitions and 
Critical Phenomena} (Clarendon Press, Oxford, 1971), p.\ 40.
\bibitem{HH}P.C. Hohenberg and B.I. Halperin, 
Rev. Mod.\ Phys. {\bf 49}, 435 and 474f (1977).
\bibitem{Need}D. Needham, T.J. McIntosh, and E. Evans, 
Biochemistry {\bf 27}, 4668 (1988).
\bibitem{Safinya}C.R. Safinya, E.B. Sirota, D. Roux, and G.S. Smith,
Phys.\ Rev.\ Lett. {\bf 62}, 1134 (1989).
\bibitem{Gui}A. Guinier, {\it Principles of X-ray Diffraction} (W.H. Freeman, 
San Francisco, 1963) ch. 9. 
\bibitem{LL}L.D. Landau and E.M. Lifschitz, {\it Statistical Physics} 
(Course of theoretical physics, v.\ 5)
(Addison-Wesley, Reading, 1958).
\bibitem{Zucker}M.J. Zuckermann, J.H. Ipsen, and O.G. Mouritsen, in
{\it Cholesterol in Membrane Models}, edited by L. Finegold (CRC Press,
Boca Raton, 1993).
\bibitem{Tan}T. Taniguchi, Phys.\ Rev.\ Lett. {\bf 76}, 4444 (1996).
\bibitem{Meakin}P. Meakin, {\it Fractals, Scaling and Growth far from 
Equilibrium} (Cambridge University Press, Cambridge, 1998)
\bibitem{Rapp99}G. Rapp, L. Finegold, and F. Richter, Biophys.\ J. {\bf 76}
(1999), in press.
\bibitem{slow}A. Aharony, in {\it Phase Transitions and Critical 
Phenomena, v.\ 6},
edited by C. Domb and M.S. Green (Academic Press, London, 1976) ch.\ 5. 
\bibitem{Pet}H.I. Petrache, N. Gouliaev, and J.F. Nagle, 
Phys.\ Rev. E {\bf 57}, 7014 (1998).
\bibitem{Sch}O. Sch\"onborn and R.C. Desai, preprint, submitted,
http://xxx.lanl.gov/abs/cond-mat/9804237.
\bibitem{Svergun}D.I. Svergun, S. Richard, M.H.J. Koch, Z. Sayers, S. Kupin,
and G. Zaccai, Proc. Natl.\ Acad.\ Sci. {\bf 95}, 2267 (1998).
\bibitem{McIntosh}T.J. McIntosh and S.A. Simon, 
Biochemistry {\bf 32}, 8374 (1993).
\bibitem{Rappolt}M. Rappolt and G. Rapp, Ber.\ Bunsenges.\ Phys.\ Chem.
{\bf 100}(1996) 1153.
\bibitem{Blume}A. Blume, in {\it Physical Properties of Biological Membranes
and their Functional Implications}, edited by C. Hidalgo 
(Plenum, London, 1988).
\end{references}
\end{document}